\documentclass[12pt]{iopart}
\usepackage{iopams}
\usepackage{graphicx}
\usepackage{xcolor}
\usepackage[breaklinks=true]{hyperref}
\hypersetup{
  colorlinks   = true, 
  urlcolor     = blue, 
  linkcolor    = blue, 
  citecolor   = magenta 
}
\usepackage{soul} 

\newcommand{\bm}[1]{\boldsymbol{#1}}

\newcommand{\eqref}[1]{(\ref{#1})}

\begin{document}
\title{Explainable Natural Language Processing with Matrix Product States}

\author{Jirawat Tangpanitanon$^{1,2, \dagger}$, Chanatip Mangkang$^{3, \dagger}$, Pradeep Bhadola$^4$, Yuichiro Minato$^5$, Dimitris G. Angelakis$^{6,7}$, Thiparat Chotibut$^{3,*}$}

\address{

$^1 $ Quantum Technology Foundation (Thailand), Bangkok, Thailand\\
$^2 $ Thailand Center of Excellence in Physics, Ministry of Higher Education, Science, Research and Innovation, Bangkok, Thailand\\
$^3 $ Chula Intelligent and Complex Systems, Department of Physics, Faculty of Science, Chulalongkorn University, Bangkok, Thailand\\
$^4 $ Centre for Theoretical Physics \& Natural Philosophy,
Nakhonsawan Studiorum for Advanced Studies,
Mahidol University, Nakhonsawan Campus, Thailand\\
$^5 $ Blueqat Inc., Japan \\
$^6 $ School of Electrical and Computer Engineering, Technical University of Crete, Chania, Greece\\
$^7 $ Centre for Quantum Technologies, National University of Singapore, Singapore}
\ead{\mailto{thiparatc@gmail.com, thiparat.c@chula.ac.th} \\
$\dagger$ Equal contributions \\
* Corresponding author}

\vspace{10pt}
\begin{indented}
\item[\today]
\end{indented}

\begin{abstract}
Despite empirical successes of recurrent neural networks (RNNs) in natural language processing (NLP), theoretical understanding of RNNs is still limited due to intrinsically complex non-linear computations. We systematically analyze RNNs' behaviors in a ubiquitous NLP task, the sentiment analysis of movie reviews, via the mapping between a class of RNNs called recurrent arithmetic circuits (RACs) and a matrix product state (MPS). Using the von-Neumann entanglement entropy (EE) as a proxy for information propagation, we show that single-layer RACs possess a maximum information propagation capacity, reflected by the saturation of the EE. Enlarging the bond dimension beyond the EE saturation threshold does not increase model prediction accuracies, so a {\it minimal model} that best estimates the data statistics can be inferred. Although the saturated EE is smaller than the maximum EE allowed by the area law, our minimal model still achieves $\sim 99\%$ training accuracies in realistic sentiment analysis data sets. Thus, low EE is not a warrant against the adoption of single-layer RACs for NLP.  Contrary to a common belief that long-range information propagation is the main source of RNNs' successes, we show that single-layer RACs harness high expressiveness from the subtle interplay between the information propagation and the word vector embeddings. Our work sheds light on the phenomenology of learning in RACs, and more generally on the explainability of RNNs for NLP, using tools from many-body quantum physics.
\end{abstract}

%
\vspace{2pc}

\section{Introduction}
The study of many-body quantum physics prompts the development of theoretical and numerical techniques to compactly represent and analyze quantum states living in an exponentially large Hilbert space. One of the most prominent examples is a compact representation of a ground state of a one-dimensional gapped quantum lattice system with local interaction as a matrix product state (MPS), which can efficiently parametrize an appropriate ground state using resources that grow only linearly with the system size \cite{Cirac_MPS_PRB}. Compact representations generalizing MPS to higher dimensional systems include projected entangled pair state (PEPS) \cite{Cirac_PEPS2004}, multiscale entanglement renormalization ansatz (MERA), which efficiently parametrize critical high-dimensional systems \cite{Vidal_2007}, and, more generally, tensor network (TN) states \cite{Orus_NatRevPhys19}. These compact state representations drastically reduce the number of parameters from exponential to at most polynomial in the system size, rendering the analysis and the simulation of many-body quantum systems computationally tractable. Dimensionality reduction has enabled insights into a broad range of many-body quantum phenomena, ranging from quantum phase transitions to topological phases of matter.

On the other hand, machine learning (ML) has also benefitted from algorithms that extract a compact representation of complex data of interests. In supervised machine learning, many algorithms efficiently convert a gigantic set of data-label pair $\{ ({\bf X}_i, l_i) \}$, where ${\bf X}_i \in \mathbb{R}^d$ is the $i^{th}$ data vector and $l_i$ is the scalar label of that data vector, into a compact representation encoding the relationship between data-label as a conditional probability $P(l | {\bf X}, {\bf \theta})$ parametrized by a set of parameters ${\bf \theta}.$ With the advent of deep learning (DL), a modern paradigm of ML that imitates computational models of biological neural networks, probabilistic relationships between data-label pairs as complex and extensive as picture-name matching, sound-text pairing, or text-speech generation, can be efficiently represented~\cite{GoodBengCour16}.  In fact, the ability of DL to extract a compact representation of complex data has fueled modern artificial intelligence technologies, including image recognition, speech recognition, and language translation, to name a few.

Since many-body quantum physics and supervised ML both benefit from a compact representation of high-dimensional mathematical objects, applying successful techniques from one discipline to the other has led to a fruitful cross-fertilization. For example, variational ansatz based on artificial neural networks offers a useful, though less interpretable, representation of complex many-body quantum states \cite{MelkoCarleo_NatPhys19, TorlaiMelko_ARCMP20, CarleoTroyer_Science17}. Automatic classification of quantum phases of matter also benefits from supervised ML approaches \cite{CarrasquillaMelko_NatPhys17}. In the opposite direction, techniques from many-body quantum physics can offer novel computational paradigms for supervised ML. References \cite{Stoudenmire_NeurIPS16, Stoudenmire_QST17, Liu_NJOP19, Cirac_IEEE20} propose tensor networks as a quantum-inspired supervised ML ansatz that can achieve high performance in image recognition tasks. Furthermore, entanglement entropy in quantum tensor networks can shed light on the information propagation in the dual artificial neural networks \cite{Levine_PRL2019,LevineYCS18}. Such duality between quantum tensor networks and artificial neural networks can help scientists scrutinize the inner working of complex neural network algorithms, providing new tools to tackle the explainability aspects of black-box DL approaches. Recent work applies quantum techniques to tackle the explainability of image recognition tasks \cite{Liu_NJOP19}, though the analysis for realistic natural language processing (NLP) tasks is still lacking.

With the goal of investigating the inner working of DL in NLP, we study the behaviors of single-layer recurrent arithmetic circuits (RACs), a class of recurrent neural networks (RNNs) that can be mapped to an MPS \cite{Levine_PRL2019}, in a ubiquitous NLP task, the sentiment analysis of movie reviews. The objective of sentiment analysis is to classify each written review into an appropriate category such as `like' or `dislike'. We show that, by using the entanglement entropy (EE) of the dual MPS as a measure of information propagation in the networks, single-layer RACs achieve highest prediction accuracies when the information propagation saturates. By saturation we mean there exists a critical model size $\chi^*$ (measured by the number of hidden neurons) such that larger models have prediction accuracies and the EE as high as those of the model with size $\chi^*$. Thus, there is a {\it minimal} single-layer RACs model that can best estimate the statistics of sentiment analysis data. The prediction accuracies are excellent though the saturated EE is below the maximum EE restricted by the area law of an MPS.

In NLP, another crucial component of successful DL models is a word embedding, a vector representation of a word that encapsulates word semantics. While the EE analysis reveals the behaviors of information propagation within single-layer RACs, it disregards the role of word embedding. In this work, we also analyze the interplay between information propagation and the word embedding. We report that, with a trainable word embedding, single-layer RACs can achieve higher prediction accuracies at a smaller EE,  compared to those of a model with a fixed embedding. As the EE drops down from its maximum value to saturation as the model size increases, the word embedding becomes more meaningful (as measured by the behavior of the cosine similarity between word vector representations.) Hence, single-layer RACs have a trade-off between achieving long-range information propagation and attaining a meaningful word embedding in sentiment analysis tasks. 

 Although a long-range correlation in an MPS is bounded above by the area law \cite{Cirac_PRL_2008}, our results demonstrate that an MPS can still serve as a useful variational ansatz in a realistic NLP task, provided the input word embedding is well designed. Recently, tensor network models for sequence modeling have been proposed \cite{Terilla_2020, Terilla_2019,Miller_Terilla_2021, Poletti_2018, Zhang_Song_2019}; however, most work focus on setting up new TN-based variational ansatz that may not have an exact mapping to the RNNs counterpart. Albeit interesting, these models do not meet our goal to scrutinize the inner working of RNNs. 

The manuscript is organized as follow. We begin with a self-contained background on a probabilistic sequence modeling and statistical language modeling in section \ref{sec:IntroMPSRNNs}. Sequence modeling in the era of RNNs and how to represent a word meaningfully as a vector are provided in sections \ref{subsec: RNNs}-\ref{subsec: embedding}. The mapping between RACs and MPS as well as the meaning of entanglement entropy as a proxy for information propagation are discussed in section \ref{subsec: MPS_RNNs}.
Section \ref{sec: results} reports numerical methods and the results of RACs performance on sentiment analysis for the IMDb data set, a standard movie review data set, revealing the information propagation capacity in RACs. Comments on RACs' behaviors and the role of word embedding are provided in details in the same section. Similar results for a smaller movie review data set (Rotten Tomatoes) are reported in the Appendix. Finally, we conclude with the discussion and outlook in section \ref{sec:discussion}.

\section{Statistical language modeling with  RNNs and MPS}\label{sec:IntroMPSRNNs}
Since language is a sequential phenomenon, in which a sequence of words (or alphabets) dictates its meaning, we first review a statistical approach to model sequences.  
 A central mathematical object for statistical sequence modeling is the joint probability distribution $P(X_{1:T})\equiv P(X_1,X_2,\dots,X_T)$ where a discrete random variable $X_t$ with $t \in \{1,\dots,T\}$ can take a value $x_t$ from a finite set $S_N$ with $N$ elements. One can regard the list of {\it correlated} random variables $X_{1:T}$ as a discrete time-series. Using Bayes' rules, the joint distribution can be factorized into the product of conditional probabilities, conditioned on the knowledge of the past, as

\begin{equation}\label{eq: chainrule}
P(X_{1:T})=P\left(X_{1}\right) P\left(X_{2} \mid X_{1}\right) P(X_3 \mid X_{1:2})\cdots P\left(X_{T} \mid X_{1:T-1}\right),
\end{equation}
where $X_{1:t}$ denotes the sequence of random variables in the first $t$ steps; i.e. $X_{1:t}\equiv \left(X_1,X_2, \dots , X_t\right)$. However, given a time series data,  inferring the conditional distribution $P\left(X_{T} \mid X_{1:T-1}\right)$ from occurrence frequency of a long data sequence can be impractical, as each realization $X_{1:T-1} =  (x_1,x_2,\dots,x_{T-1})$ typically occurs with a relative frequency $\sim N^{1-T}$ (assuming $X_t$ is uniformly distributed over $S_N$), which is exponentially small in $T$. 

One may assume a short temporal correlation in the sequence, so that the conditional probabilities depend only on the $n$ previous steps  
\begin{equation} \label{eq: n-gram}
P(X_{T} \mid X_{1:T-1}) \approx P(X_T \mid X_{T-n: T-1}).
\end{equation}
For a small $n$, a realization $X_{T-n:T-1} =  (x_{T-n},\dots,x_{T-1})$ now occurs with a non-negligible frequency $\sim N^{-n} \gg N^{1-T}$, rendering the estimation of \eqref{eq: n-gram} manageable. For $n=1$, \eqref{eq: n-gram} is the familiar Markov assumption, which gives a Markovian approximation of a stochastic process $X_{1:T}$ in \eqref{eq: chainrule} .

In the context of natural language processing (NLP), a probabilistic model that prescribes probabilities to sequences of words (or alphabets) is called a {\it language model} \cite{Jurafsky_SpeechLanguage2009}. Predicting the next word (or alphabet), given a sequence of previous words (or alphabets) is one important example with myriad applications. A  model that predicts the next word based only on the last $n-1$ words according to \eqref{eq: n-gram} is called an {\it n-gram} language model. 

However, even with a small $n = 4$, constructing a {\it 4-gram} model from a gigantic text, such as all the Wikipedia's english articles, can be impractical. Consider a random variable $X_t$ in \eqref{eq: chainrule} which now takes a realization as a word $w_t$ from $S_N$, a dictionary with $N$ words. Oxford English dictionary contains $N = 171,476 \approx 10^5$ English words that are currently used \cite{oxford_dict}. In this case, the frequency of occurrence of a sequence of 4 words $(w_{t-4},w_{t-3},w_{t-2},w_{t-1})$ can be vanishingly small $\sim N^{-4} \approx \left(10^5\right)^{-4} = 10^{-20}$, rendering the estimation of $P(X_t \mid X_{t-4: t-1})$ impractical. In addition, accurate prediction of the next word often depends on the words that appear in the far past. 
For instance, the prediction accounting for the subject-verb agreement in 
``This example that we demonstrate for you \_\_\_ " can be syntactically wrong if the model retains only the last few words, whose prediction would be ``are". It requires a 7-$gram$ model to correctly predict ``is" in this example. Therefore, to construct a useful language model, one needs to devise a computational approach that can encapsulate a long-range correlation in a sequence of words, while also circumventing the sparsity of a long sequence problem. This can be achieved with recurrent neural networks (RNNs), which we now discuss.

\subsection{Statistical language modelling with RNNs}\label{subsec: RNNs}


\begin{figure}\includegraphics[width = 0.9\textwidth]{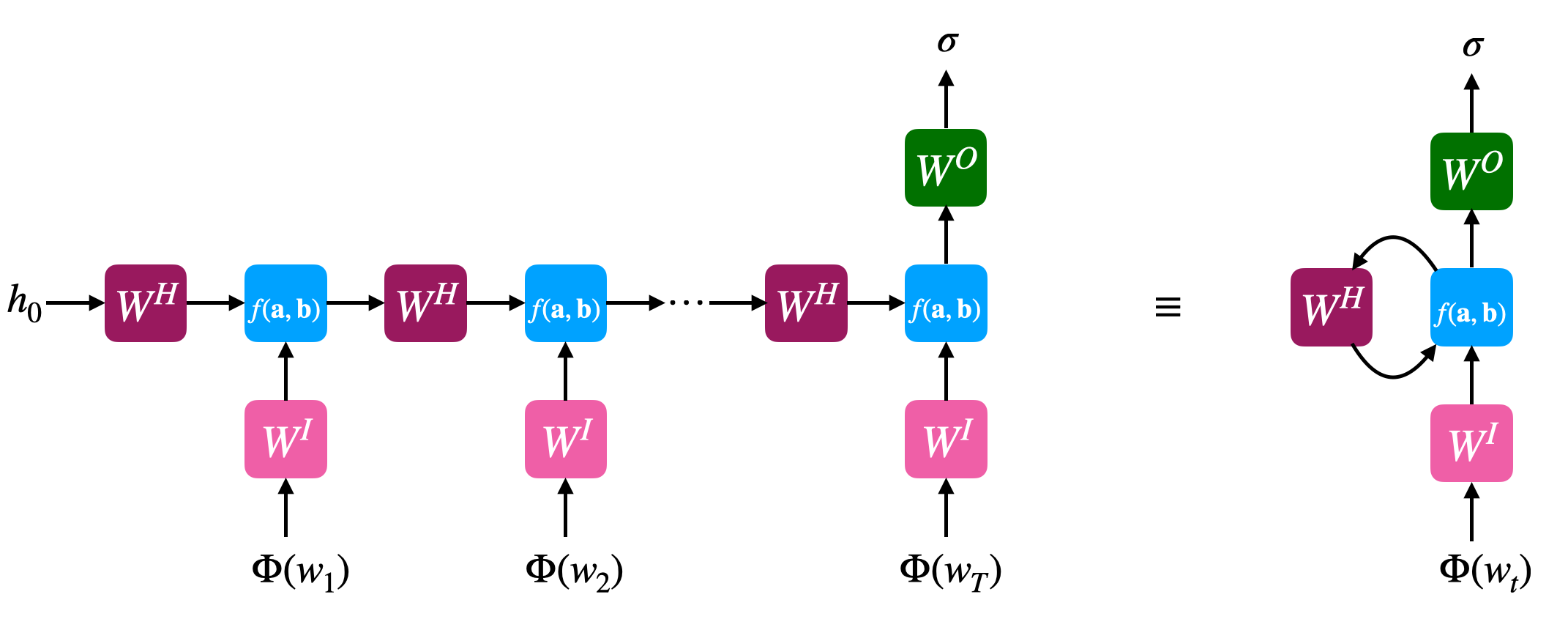}
\caption{A schematic for the sentiment analysis (binary classification) of a word sequence $w_{1:T}$ performed by a vanilla recurrent neural network (RNN), which outputs the probability that the sequence has a positive sentiment $\sigma \in [0,1]$. Each word $w_t$ is embedded as a vector $\bm{\Phi}(w_t) \in \mathbb{R}^{d_I} $, with $d_I$ of around 300-500 for other large language modeling tasks \cite{word2vec, Glove}. The recurrent computation is iterated from a dynamical system $\bm{h}_{t}= f\left(W^{\mathrm{H}} \bm{h}_{t-1}, W^{\mathrm{I}} \bm{\Phi}\left(w_t\right)\right) \in \mathbb{R}^{d_H}$, with some non-linear map (activation function) $f: \mathbb{R}^{d_H} \times \mathbb{R}^{d_I} \rightarrow  \mathbb{R}^{d_H} $. In the last time step $t=T$, when sentiment classification is performed, one computes the sigmoid function $\sigma \left(W^O \boldsymbol h_T + b^O\right)$ which assigns the probability that the sequence $w_{1:T}$ has a positive sentiment.} 
\label{fig: RNNs Architecture}
\end{figure}

Rather than conditioning the prediction task on a window of size $n$ as in an {\it n-gram} language model, RNNs allow conditioning the prediction on {\it all} previous words that appear in a text, approximating \eqref{eq: chainrule} while requiring only finite computational resources. RNNs can also perform the next item prediction as it is widely used to estimate the conditional probability $P(X_T | X_{1:T-1})$; however, one ubiquitous yet simpler task in NLP is to estimate $P(\sigma | w_{1:T})$, where $\sigma$ is a discrete quantity that characterizes a sequence of words $w_{1:T}$ of length $T$. We will focus on characterizing the sentiment of written sentences, a task termed {\it sentiment analysis} in NLP, in which $\sigma$ is a {\it binary} variable that takes a value 0 if a given sequence has a negative sentiment, and a value 1 if a given sequence has a positive sentiment. This task can be used, for example, to automatically rate product reviews or analyze news sentiment. 

We now describe the simplest (Elman's/vanilla) RNN that is typically adopted to approximate conditional probabilities. Figure \ref{fig: RNNs Architecture} shows a recurrent computational unit that estimates $P(\sigma|w_{1:T})$. At every time step, except the first and the last, this recurrent computational unit computes from an input $\boldsymbol{\Phi} (w_t)$ of dimension $d_I$ and a {\it hidden} or {\it latent} vector $\boldsymbol h_{t-1}$ of dimension $d_H$ a non-linear output function $f$, called {\it output hidden vector} $\boldsymbol h_{t}$ of dimension $d_H$, 

\begin{equation}\label{eq: activation}
	\boldsymbol h_t \equiv f(W^I\boldsymbol{\Phi}(w_t), W^H \boldsymbol h_{t-1}, \boldsymbol b).
\end{equation}
Here $W^I$ is the input weight matrix with dimension $d_H \times d_I$ that aggregate signals from the input vector $\boldsymbol{\Phi}(w_t)$ , $W^H$ is the weight matrix aggregating the signals from the hidden vector whose dimension is $d_H \times d_H$, and $\boldsymbol b$ is the so called bias with dimension $d_H$. The non-linear {\it activation function} $f$ imitates the behaviors of biological neurons, such that the weighted input and the weighted hidden vector are summed together (mimicking aggregation of potentials), while the bias (representing the background neuron's potential) is added to the aggregated weighted sum. In Elman's RNNs, each component $\boldsymbol s_t^{(i)}$ of the aggregated sum $\boldsymbol s_t = W^I\boldsymbol{\Phi}(w_t) + W^H \boldsymbol h_{t-1} + \boldsymbol b$ represents a total potential each hidden neuron $i \in \{1,\dots,d_H\}$ experiences, which triggers each hidden neuron to be activated and outputs a corresponding component of the hidden vector into the next time step as $\boldsymbol h_{t}^{(i)} =f(\boldsymbol s_t^{(i)}).$ The notation in the last equality and \eqref{eq: activation} signifies that the {\it same} activation function $f$ is applied identically to every hidden neuron. Standard non-linear activation function $f$ motivated by neurobiology is a sigmoid function or tanh function, whereas modern machine learning typically employs rectified linear unit (ReLU) defined by $f(x) = \textrm{Max}\{0, x\}$ and its variants \cite{GoodBengCour16,MEHTA20191}.

With \eqref{eq: activation} as a computational building block, one can iterate the computation recursively taking into account all the inputs in the sequence $\boldsymbol \Phi_{1:T} \equiv \left(\boldsymbol \Phi(w_1), \dots,\boldsymbol \Phi(w_T)\right)$ of size $T$, provided a hidden vector $\boldsymbol h_0$ was initialized. Due to the recursive structure, it's plausible that information in the far past can influence the output vector at the last step $\boldsymbol h_{T}$. This manifestation of long-term temporal dependencies  through a recursive computation circumvents the problem of an astronomical number of parameters needed to model a long sequence encountered in the previous section. Here one only requires to store the bias vector $\boldsymbol b$ of dimension $d_H$, $W^I$ of dimension $d_H \times d_I$, and $W^H$ of dimension $d_H \times d_H$, all of which are independent of the sequence length $T$.

The simplest sentiment analysis task, which we focus on, is a binary classification task where there are only 2 sentiments $\sigma \in \{0,1\}$. In such case, the final hidden vector will be passed to the classification neuron with the output weight $W^O$ whose dimension is $1 \times d_H$ together with the added scalar bias $b^O$ as the aggregated signal of the classification neuron $s_{O} = W^O \boldsymbol h_{T} + b^O$ before the classification neuron predicts a number $\hat \sigma_{\theta} \in [0,1]$ computed from the sigmoid activation function
\begin{equation}\label{eq: sigmoid}
	\hat \sigma_{\theta} \equiv \frac{1}{1+\exp(-s_O)},
\end{equation} 
where we denote the set of all parameters in this RNN that influences the value of this last neuron as $\theta \equiv \{\boldsymbol{b}, W^I, W^H, W^O, b^O\}$. 

To train the model, one adjusts parameters $\theta \equiv \{\boldsymbol{b}, W^I, W^H, W^O, b^O\}$ to minimize the cost (loss) function $C$ which accumulates the amount of mismatches between the true sentiment $\sigma(\boldsymbol w_m)$ associated with the $m^{th}$ sequence $\boldsymbol w_m \equiv w_{1:T,m} = \left(w_1,w_2,\dots,w_T \right)_m$ and the RNNs' sentiment prediction $\hat \sigma_{\theta}(\boldsymbol w_m)$, for all sequences in the training sample $m \in \{1,\dots,M\}$. For a binary classification task with the probabilistic prediction given by \eqref{eq: sigmoid}, the cost function is typically taken as the binary cross-entropy
\begin{equation}\label{eq: cross-entropy_cost}
	C \equiv \frac{1}{M}\sum_{m=1}^M \Big( \sigma(\boldsymbol w_m) \log\left[\hat\sigma_{\theta}(\boldsymbol w_m)\right] + \left( 1- \sigma(\boldsymbol w_m) \right)\log\left[1-\hat\sigma_{\theta}(\boldsymbol w_m)\right] \Big).
\end{equation}
In a movie review task, for example, $M$ can be the number of written reviews with predetermined sentiments from $M$ different reviewers that encapsulates a reasonable relationship between word sequences and their associated sentiments.  

Note that minimizing the cross-entropy $C \equiv H_{RNN_{\theta}}\left(P\right)$ between the empirical distribution $P(\sigma|w_{1:T})$ constructed from the training data and the distribution predicted by the RNN parametrized by $\theta$, denoted by $RNN_{\theta}(\sigma | w_{1:T})$, is equivalent to minimizing the KL divergence $D_{KL}\left(P\mid\mid RNN_{\theta}\right)$ \cite{murphy2013machine,bishop2007}. Since the KL divergence reflects the dissimilarity between the two distributions, the optimization (minimization) procedure of the cost function \eqref{eq: cross-entropy_cost} would search for a vanilla RNN parametrized by $\theta^*$ that estimates well the empirical distribution $P(\sigma|w_{1:T}).$ Provided the training data is properly curated and the optimization procedure (e.g., gradient methods and their modern variants \cite{MEHTA20191, GoodBengCour16}) is reliable, one shall arrive at a reasonable statistical relationship between a {\it long} sequence of words and its associated sentiment parametrized by an RNN with a {\it finite} number of parameters $\theta^*$. In other words, 
\begin{equation}\label{eq: sentiment_conditional_prob}
	P(\sigma|w_{1:T}) \approx RNN_{\theta^*}(\sigma|w_{1:T}).
\end{equation} 
This is the main philosophy behind statistical language modeling using recurrent neural networks.

\subsection{On the word vector embedding $\boldsymbol \Phi$}\label{subsec: embedding}

Suppose one randomly assigns or `tokenizes' each word with a unique integer $w_i\in \{1,\dots,N\}$, where $1\le i\le N$ with $N$ being the size of the dictionary. Then each written review is represented by a sequence of integers $w_{1:T}=(w_1,...,w_T)$. Here, the length of each review is forced to be $T$, by padding $0$'s at the beginning of the review if its length is less than $T$, or by selecting only the first $T$ words if its length is greater than $T$. For example, for $T=6$, the sentence `Physics is beautiful' can be encoded as $w_{1:T}=(0,0,0,532,3,46)$, where `Physics'$=532$, `is'$=3$, and `beautiful'$=46$. The tokenization process, however, artificially introduces the notion of distance between two words that does not encode word semantics. 

How shall one mathematically represent words so that their semantics are encoded?  A widely-adopted solution is to embed a word $w$ as a vector $\boldsymbol \Phi(w) \in \mathbb{R}^{d_I}$. By representing a word as a vector embedded in $d_I$ dimensions, words with similar meanings that co-occur frequently in the same context can be assigned unique vectors such that their pairwise Euclidean distance are small. Also, a negative {\it cosine similarity} of the embeddings of the two words $w_a, w_b$  computed from
\begin{equation}\label{eq: cosinesim}
	\textrm{sim} \left( \boldsymbol \Phi (w_a), \boldsymbol \Phi(w_b) \right) = \frac{\boldsymbol \Phi (w_a)}{||\boldsymbol \Phi (w_a)||_2}\cdot \frac{\boldsymbol \Phi (w_b)}{||\boldsymbol \Phi (w_b)||_2}
\end{equation}
can signify that $w_a$ and $w_b$ rarely co-occur in the same context, and hence could have opposite meanings.
 
There are a few methods to numerically obtain an embedding $ \Phi$ that effectively represents word semantics \cite{word2vec, Glove, peters-etal-2018-deep}. A simple yet classic Word2vec method \cite{word2vec}, which is also adopted in our numerical experiments, is to assign the embedding function $\Phi$ as a matrix of size $d_I\times N$, so that the $i^{\rm th}$ column of $\Phi$ corresponds to the word vector $\boldsymbol \Phi(w_i)$ of the word $w_i$.  The embedding dimension $d_I$ is a hyper-parameter that can be tuned to best suit the problem. The matrix elements in $\Phi$ are treated as variational parameters to be optimized along with the optimization of the RNN for a language modeling task of interests. For example, to perform a sentiment analysis using a vanilla RNN without knowing {\it a priori} the embedding matrix, one would add the matrix elements of $\Phi$ into the trainable parameters $\tilde \theta \equiv \{ \theta, \Phi \} = \{\boldsymbol{b}, W^I, W^H, W^O, b^O, \Phi\}.$ In this way, training the RNN according to section \ref{subsec: RNNs} will not only yield the network parameters, but also the word vector embedding. With a sufficiently large and well curated training data set, one expects that the embedding matrix $\Phi$ would effectively encapsulate word semantics in the dictionary of interests. 

Despite the empirical success of statistical language modeling using vanilla RNNs together with the well-trained word embedding as explained above, highly-nonlinear iterations of \eqref{eq: activation} by standard activation functions render the analysis of how RNNs approximate empirical sequence distributions very challenging. In the following, we review recent attempts to analyze the expressiveness of RNNs (i.e. the set of function that can be effectively parametrized by RNNs) with a specific activation function, through the mapping to their dual the tensor network counterparts. 

\subsection{Recurrent Arithmetic Circuit (RAC) and the mapping to Matrix Product State (MPS)}\label{subsec: MPS_RNNs}
\begin{figure}
\includegraphics[width = 0.8\textwidth]{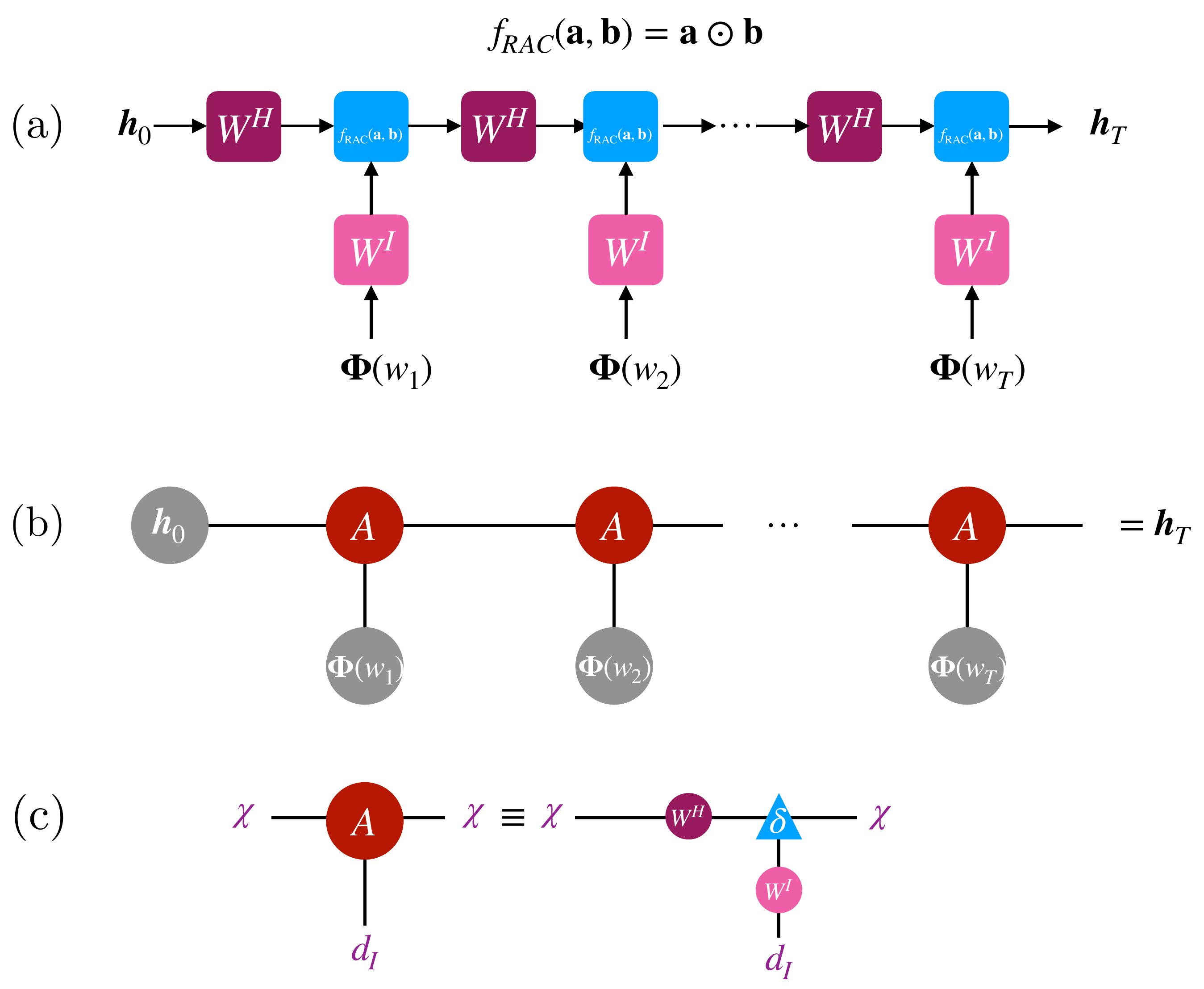}
\caption{The graphical representation of the mapping between a single-layer RACs for sentiment analysis task (a) to the dual MPS (b). As a fundamental building block, the translational invariant MPS (without the contraction by boundary vector $\boldsymbol h_0$) consists of the rank-3 tensor $A^{s_t}_{\alpha_{t} \alpha_{t-1}} \equiv \sum_{\tilde \alpha_{t-1}, \tilde s_t =1}^{d_H} W^H_{\tilde{\alpha}_{t-1}\alpha_{t-1}}\delta_{\alpha_{t} \tilde{\alpha}_{t-1} \tilde{s}_t}W^I_{\tilde{s}_t s_t},$ where the triangle in (c) represents the tensor of rank 3 defined by $\delta_{jkl}$ which is equal to 1 if $j=k=l$ and is 0 otherwise. The structure of the building block in (c) arises from the Hadamard product imposed by RAC activation function in (a). Here we denote $\chi$ as the bond dimension of the MPS, which is equal to $d_H$, the number of hidden units of RACs  in (a). The vertical bond in (c) has the dimension $d_I$, identical to that of the word vector embedding $\boldsymbol \Phi$.} 
\label{fig: MPS_RNN map}
\end{figure}

Consider the activation function defined by the Hadamard product
\begin{equation}
f_{RAC}(\boldsymbol a, \boldsymbol b)	= \boldsymbol a \odot \boldsymbol b, 
\end{equation}
which is the element-wise multiplication $f^{(i)}_{RAC}(\boldsymbol a, \boldsymbol b) \equiv \boldsymbol a^{(i)}\cdot\boldsymbol b^{(i)}$.
Recurrent neural networks with RAC activation function, known as Recurrent Arithmetic Circuits (RACs), have recently received increasing attention and share computational paradigm similar to the Multiplicative Recurrent Neural Networks \cite{Wu_Bengio_NIPS2016, Sutskever_ICML, Poon_IEEE, Bengio_NIPS2011}. More importantly, references \cite{LevineYCS18, Levine_PRL2019} show that a single-layer RAC can be mapped to the dual Matrix Product State (MPS), taking the inspiration from the Tensor Train (TT) decomposition of \cite{Oseledets_SIAM_TT}. By studying RACs, the analysis of learning in RNNs for temporal data can thus be performed from many-body quantum physics perspectives. For instance, one can compute the entanglement entropy of the dual MPS, which is a measure of the amount of temporal correlation that can be supported by the network \cite{LevineYCS18, Levine_PRL2019}. The larger the entanglement entropy means that the output of network computation crucially depends on the temporal data in the further past, enabling the network to have a longer-range memory. 

The tensor network diagrams in figure \ref{fig: MPS_RNN map} summarize the equivalence between the computation of the standard RNNs-based sentiment analysis with RAC activation function and that of the dual matrix product state. By defining the tensor of rank 3 of the form 
\begin{equation}\label{eq: A_tensor}
A^{s_t}_{\alpha_{t} \alpha_{t-1}} \equiv \sum_{\tilde \alpha_{t-1}, \tilde s_t =1}^{d_H} W^H_{\tilde{\alpha}_{t-1}\alpha_{t-1}}\delta_{\alpha_{t} \tilde{\alpha}_{t-1} \tilde{s}_t}W^I_{\tilde{s}_t s_t},	
\end{equation}
where $\delta_{jkl}$ is 1 if $j=k=l$ and is 0 otherwise, the state evolution by one time step can be computed by the tensor contraction between the hidden vector $\boldsymbol h_{t-1}$, the tensor $A^{s_t}_{\alpha_{t} \alpha_{t-1}}$, and the input word vector $\boldsymbol{\Phi}(w_t)$, resulting in the tensor of rank 1 describing the hidden vector of the next time step whose component $\alpha_{t}$ is given by
\begin{eqnarray*}
	\boldsymbol{h}_{t}^{(\alpha_{t})} = \sum_{\alpha_{t-1}=1}^{d_H}\sum_{s_t=1}^{d_I} \boldsymbol{h}_{t-1}^{(\alpha_{t-1})}A^{s_t}_{\alpha_{t} \alpha_{t-1}} \boldsymbol{\Phi}^{(s_t)}(w_t)\\
\hspace{1.3cm}=\sum_{\tilde \alpha_{t-1}, \tilde s_t =1}^{d_H} \sum_{\alpha_{t-1}=1}^{d_H}\sum_{s_t=1}^{d_I} \left(W^H_{\tilde{\alpha}_{t-1}\alpha_{t-1}}\boldsymbol{h}_{t-1}^{(\alpha_{t-1})}\right)\delta_{\alpha_{t} \tilde{\alpha}_{t-1} \tilde{s}_t}\left(W^I_{\tilde{s}_t s_t}\boldsymbol{\Phi}^{(s_t)}(w_t)\right)\\
\hspace{1.3cm}= \left(W^H \boldsymbol{h}_{t-1}\right)^{(\alpha_{t})}\cdot \left( W^I \boldsymbol{\Phi}(w_t)\right)^{(\alpha_{t})}\\
\hspace{1.3cm}=f^{(\alpha_{t})}_{RAC}(W^H \boldsymbol{h}_{t-1},W^I \boldsymbol{\Phi}(w_t)).
\end{eqnarray*}

Therefore, given a sequence $\left(\boldsymbol \Phi(w_1), \dots, \boldsymbol \Phi(w_T)\right)$ and the initialization of the hidden vector $\boldsymbol h_0$ with dimension $d_H$, the output hidden vector at time $T$ can be computed from the contraction between the translational invariant MPS  
\begin{equation}\label{eqn: MPS}
	\Psi^{s_{T}\dots  s_2 s_1}_{\alpha_{T}\alpha_{0}} \equiv \sum_{\alpha_1, \alpha_2,\dots, \alpha_{T-1}=1}^{d_H}A^{s_T}_{\alpha_{T} \alpha_{T-1}}\cdots A^{s_2}_{\alpha_2 \alpha_1} A^{s_1}_{\alpha_1 \alpha_0},
\end{equation} 
the tensor of rank $T$ constructed from the input sequence 
\begin{equation}\label{eqn: tensor_inputseq}
	\Phi^{s_T \dots s_2 s_1} \equiv \boldsymbol{\Phi}^{(s_T)}(w_T)\cdots \boldsymbol{\Phi}^{(s_2)}(w_2)\boldsymbol{\Phi}^{(s_1)}(w_1) ,
\end{equation}
and the initial hidden vector $\boldsymbol h_0$ as follows
\begin{equation}\label{eqn: MPS_contraction}
	\boldsymbol{h}^{(\alpha_{T})}_{T} = \sum_{\alpha_0=1}^{d_H}\sum_{s_1,s_2,\dots,s_T=1}^{d_I}\left(\Phi^{s_T\dots s_2 s_1}\right)\left(\Psi^{s_T \dots s_2s_1}_{\alpha_{T}\alpha_{0}}\right)\boldsymbol{h}_0^{(\alpha_0)}.
\end{equation}
The last equality is compactly represented by the standard tensor network graphical notation as shown in figure \ref{fig: MPS_RNN map}(b), whose building block is the tensor $A$ of \eqref{eq: A_tensor} represented graphically in figure \ref{fig: MPS_RNN map}(c).  For sentiment analysis using binary classification, the final contraction \eqref{eqn: MPS_contraction} will then be used to compute the probability that the input sequence $w_{1:T}$ has a positive sentiment through the usual sigmoid activation function as in \eqref{eq: sigmoid}. Note that, in many-body quantum physics language, the dimension of the hidden unit $d_H$ is in fact the {\it bond dimension} $\chi \equiv d_H$ of the MPS.

\subsection{Entanglement entropy of the MPS as a proxy for information propagation in RAC}\label{subsec: EE_MPS}

Since the fundamental building block of the computation is the translational invariant MPS, we can compute the entanglement entropy (EE) by partitioning the MPS into two subsystems through the standard Schmidt-decomposition, and compute the resulting von-Neumann entropy \cite{Ekert_Knight95}. However, the MPS in \eqref{eqn: MPS} still has an open boundary. To make the boundary close and properly compute the EE, one needs to contract the indices $\alpha_0,$ and $\alpha_T$ by vectors of dimension $\chi = d_H$. In the limit $T \gg 1$, this choice of vectors should not significantly affect the EE if the partition is made at half of the chain. The details on an appropriate choice of vectors for contraction to close the boundary in our numerical experiments will be discussed in the following section. Suppose now that the contraction has been properly made and the MPS with a close boundary is given by $\tilde \Psi^{s_T\dots s_2 s_1},$
then the corresponding quantum state of the MPS is 
\begin{equation}
	| \tilde \Psi \rangle_{MPS} = \sum_{s_1,\dots, s_T = 1}^{d_I} \tilde \Psi^{s_T\dots s_2 s_1} |s_T\rangle \otimes \dots \otimes | s_2 \rangle \otimes |s_1\rangle,
\end{equation}
which has the Schmidt decomposition (singular value decomposition) for the bipartition at the $\lceil T/2 \rceil^{th}$ bond into the left and right sectors as
\begin{equation}
	|\tilde \Psi\rangle_{M P S}=\sum_{i=1}^{r} \lambda_{i}\left|\phi_{i}^{L}\right\rangle \otimes\left|\phi_{i}^{R}\right\rangle,
\end{equation} 
where the Schmidt coefficients $\lambda_i$'s are the real, non-negative singular values satisfying $\sum_{i=1}^r \lambda_i^2 = 1$, and $r$ is the Schmidt rank (Schmidt number). The Schmidt rank $r$ is 1 only for a product state and is greater than $1$ when a state has the two subsystems that are entangled. 

The von-Neumann (entanglement) entropy is a well-defined measure of entanglement between the two subsystems and can be calculated as
\begin{equation}\label{eqn: EE}
	S=-\sum_{i=1}^{r} \lambda_{i}^{2} \log _{2} \lambda_{i}^{2}.
\end{equation} 
Importantly, this entanglement entropy, when translated into the recurrent neural network language, can quantify the amount of temporal correlation between the signal in the earlier times $\{\boldsymbol h_1, \dots, \boldsymbol h_{\lceil T/2 \rceil}  \}$ and the signal in the later times $\{\boldsymbol h_{\lceil T/2 \rceil+1}, \dots, \boldsymbol h_T  \}$, also known as {\it Start-End separation rank} \cite{levine2018benefits, Levine_PRL2019}. If the entanglement entropy is zero, the signals in the earlier and the later times are statistically independent. The prediction task from models with vanishing EE thus has a short-term memory, neglecting the knowledge in the past $t < \lceil T/2 \rceil$. One then would expect the models with larger EE to be more desirable in encapsulating long-range sequence correlations. We shall then intuitively interpret the EE computed from \eqref{eqn: EE} as {\it the proxy for  information propagation} in the RACs networks.   RACs that possess low EE might have a low expressiveness (high bias in statistical learning theory framework), and thus are unable to efficiently approximate data distribution with long-range statistical correlations. 

It's well known that an MPS obeys the area law of entanglement entropy, which constrains the upper bound on EE as $S = O(\log_2(\chi))$ \cite{Eisert_RMP_AreaLaw}. In fact, the state with the maximum entropy in \eqref{eqn: EE}   is attained with the value $\log_2(\chi)$ when all the Schmidt coefficients are identically $1/\sqrt{\chi}$ with the Schmidt rank $r = \chi$.\footnote{The discrete distribution $P_i$ that maximizes the Shannon's entropy $-\sum_{i=1}^m P_i \log_2 P_i$ is the uniform distribution $P_i = 1/m$.} Since the upper bound is independent of the system size $T$, temporal data with  long-range statistical correlation might not be efficiently approximated by an MPS (or, equivalently, single-layer RACs) variational ansatz. This result seems to warrant a no-go statement for using MPS to model sequential data with long-range correlation. Alternative models that can incorporate long-range correlation, such as Deep RACs, have been theoretically analyzed, though no experimental results on these network performance on realistic temporal data sets have been reported \cite{Levine_PRL2019, LevineYCS18, levine2018benefits}. 

However, thus far, the analysis on the expressive power of single-layer RACs concerns only that of the recurrent units, not of the combined system that includes a representation $\Phi$ of the input embedding. In practice, even in simple RNNs, incorporating trainable word embedding function $\Phi$ into the model can tremendously increase the prediction accuracy. In the following section, we shall investigate, in realistic sequence modeling settings, whether low EE of models alone suffices to enforce a no-go theorem for such models. The answer is an affirmative no, and single-layer RACs are still useful in realistic sequence modeling tasks.

%
%
%

\section{Sentiment analysis by single-layer RACs with an entanglement entropy below the area law: numerical experiments}\label{sec: results}

In this section, we first provide the details of our numerical experiments to analyze the behaviors of single-layer RACs for sentiment analysis in realistic movie reviews data sets. Then, we discuss the importance of additive biases in RAC activation function, and elucidate how to convert RACs with additive biases into MPS for the purpose of entanglement entropy analysis.  We then report the behaviors of single-layer RACs together with their entanglement entropy. First, we show that when a pre-trained word vector embedding $\boldsymbol \Phi$ is fixed, the prediction accuracies strongly correlate with the amount of information propagation within RACs as measured by the entanglement entropy. Interestingly, the high prediction accuracies saturate when the entanglement entropy saturates, enabling one to determine the minimal model (model with the smallest bond dimension $\chi^*$ that saturates the entanglement entropy) that can best approximate the statistics of sequential data. This entanglement entropy saturation is a reflection of  the convergence of entanglement spectrum to the limiting entanglement spectrum that we numerically report. Second, when the embedding layer is trained along with RACs, there is an intriguing interplay between RACs and the embedding layer such that, even when the entanglement entropy drops, the prediction accuracy is boosted. Contrary to a common belief that long-range information propagation in the network is the main source of RNN's expressiveness, we show that, when the bond dimension is large, RACs harness its high expressiveness from meaningful word embeddings.
  
	
\subsection{Details of the numerical experiments}\label{subsec:sentiment}

In the main text, we use the IMDb movies and critic reviews data set, which is one of the standard data sets for sentiment analysis using binary classification \cite{IBDb-dataset}. The training set and the test set contain $M=$ 40,000 and 10,000 different samples respectively. Both sets are approximately balanced: the ratio of positive to negative reviews in the training and the test set are given by, respectively, 20,027:19,973 and 4,913:5,027.  The length of each review is set to $T=50$ and the dictionary size is $N=10,000$. We also perform sentiment analysis on the Rotten Tomatoes (RT) data set using the same methodology which leads to similar conclusions as the ones presented in this section. The details and the results for RT data sets are shown in the Appendix.

To train the model, we implement single-layer RACs using Keras \cite{chollet2015keras} which is a high-level API of TensorFlow. Batch training is deployed with 200 epochs with the batch size of 128. An early stopping is applied to terminate the training process if the change in the cost function after 4 epochs is smaller than  0.001. The cost function is optimized using Adam optimizer. The optimization process is repeated 50 times, each with a random initialization of the variational parameters, and the averaged prediction accuracies for the training and the test data set are obtained for each number of hidden neurons $d_H$.

\subsection{Entanglement entropy of single-layer RACs with additive biases}
It is important to note that for RACs not to suffer from the vanishing or exploding gradient problem during model training\footnote{Since RACs iteratively multiply signals, backpropagation during gradient computation can lead to the iterated product of very small numbers or very large numbers for poorly initialized training parameters, leading to vanishing or exploding gradients problem respectively. Adding a trainable bias is a way to control the scale of multiplicative iteration and help mitigate the vanishing or exploding gradients problem.}, we found that it is crucial to add  trainable bias vectors $\boldsymbol b_H, \boldsymbol b_I \in \mathbb{R}^{\chi}$ to the aggregated inputs of the RAC activation function. In particular, to achieve model trainability in practice requires the time evolution of the form $	\boldsymbol h_t \equiv f_{RAC}(W^I\boldsymbol{\Phi}(w_t)+\boldsymbol b_I, W^H \boldsymbol h_{t-1} + \boldsymbol b_H).$ Fortunately, recasting the recurrent computation with additive bias vectors as the MPS structure only requires a minor modification to the prescriptions in the previous section, which we now discuss. 

\begin{figure}
\includegraphics[width = 0.8\textwidth]{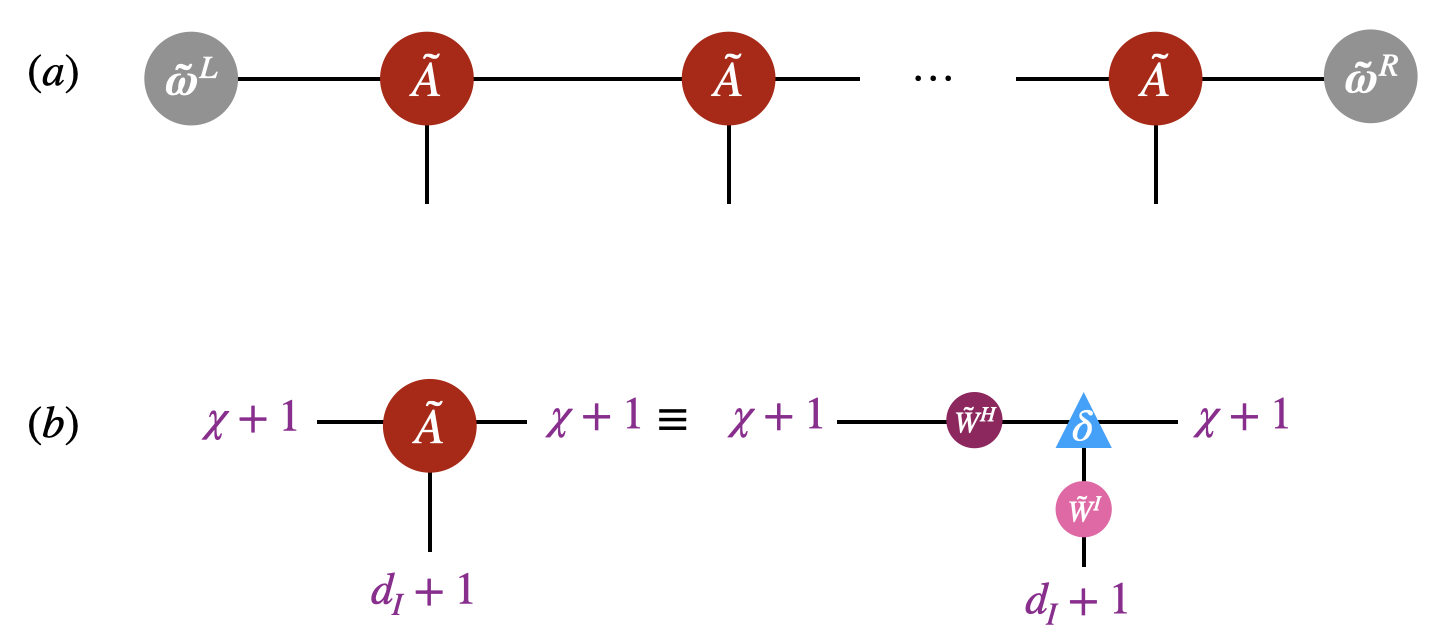}
\caption{(a) The modified translational invariant MPS in a close boundary for calculating the entanglement entropy of RACs with additive biases $\boldsymbol b_I, \boldsymbol b_H$. (b) The tensor $\tilde A^{s_t}_{\alpha_{t} \alpha_{t-1}}$ is defined according to \eqref{eq: redef_A_tensor}, which consists of the contraction between $\delta_{jkl}$ and the modified weight matrices $\tilde W^H, \tilde W^I$ of \eqref{eq: redef_W}.}
\label{fig: mps_bias_closeboundary}
\end{figure}

Define the new input and hidden weight matrices as 

\begin{equation}\label{eq: redef_W}
	\tilde W^I \equiv 
\underbrace{\left[\begin{array}{c|c}
W^I & \boldsymbol b_I \\
\hline 0 \ldots 0 & 1
\end{array}\right]}_{d_I+1}\Big \} \chi+1, \ \ \ \ 
\tilde W^H \equiv 
\underbrace{\left[\begin{array}{c|c}
W^H & \boldsymbol b_H \\
\hline 0 \ldots 0 & 1
\end{array}\right]}_{\chi+1}\Big \} \chi+1.
\end{equation}
Define also the new word vector embedding and the new hidden vector 
\begin{equation}
\tilde{\boldsymbol \Phi}(w_t)\equiv \left[ 
\begin{array}{c}
\boldsymbol \Phi(w_t) \\
1\\
\end{array}	
\right]
, \ \ \ \ 
\tilde{\boldsymbol h}_t \equiv \left[ 
\begin{array}{c}
\boldsymbol h_t \\
1\\
\end{array}	
\right].
\end{equation}
These definitions give

\begin{equation}
\tilde W^I \tilde{\boldsymbol \Phi}(w_t) =  	
\left[ 
\begin{array}{c}
W^I\boldsymbol \Phi (w_t) + \boldsymbol b_I \\
1\\
\end{array}	
\right], \ \ \ \ 
\tilde W^H \tilde{\boldsymbol h}_t =  	
\left[ 
\begin{array}{c}
W^H\boldsymbol h_t + \boldsymbol b_H \\
1\\
\end{array}	
\right].
\end{equation}
Therefore, 
\begin{equation}
\tilde{\boldsymbol h}_t = 
\left(\tilde W^H \tilde{\boldsymbol h}_{t-1}\right)\odot\left(\tilde W^I \tilde{\boldsymbol \Phi}(w_t) \right) \\
=
\left[ 
\begin{array}{c}
f_{RAC}(W^H\boldsymbol h_{t-1} + \boldsymbol b_H, W^I\boldsymbol \Phi (w_t) + \boldsymbol b_I) \\
1\\
\end{array}	
\right].
\end{equation}
The last equality states that the time evolution from RAC with the bias vectors of the original problem can be encoded into the time evolution from standard RAC (without additive biases) in one higher dimension, resulting in the translational invariant MPS with the following tensor as a building block
\begin{equation}\label{eq: redef_A_tensor}
\tilde A^{s_t}_{\alpha_{t} \alpha_{t-1}} \equiv \sum_{\tilde \alpha_{t-1}, \tilde s_t =1}^{\chi+1} \tilde W^H_{\tilde{\alpha}_{t-1}\alpha_{t-1}}\delta_{\alpha_{t} \tilde{\alpha}_{t-1} \tilde{s}_t}\tilde W^I_{\tilde{s}_t s_t},	
\end{equation}
where the input index $s_t$  now takes the value from $\{1,\dots,\chi+1\}$. 

After we obtain all the variational parameters including $\boldsymbol b_I, \boldsymbol b_H$ at the end of a training procedure with Adam optimizer, $\tilde A^{s_t}_{\alpha_{t} \alpha_{t-1}}$ that defines the MPS/TT with an open boundary can be constructed. The entanglement entropy is computed according to section \ref{subsec: EE_MPS}, where we close the left and the right boundary by the contraction with the boundary vectors $\boldsymbol{\tilde \omega}^L, \boldsymbol{\tilde\omega}^R \in \mathbb{R}^{\chi+1}$ defined by
\begin{equation}
\boldsymbol{\tilde\omega}^L \equiv \tilde{\boldsymbol h}_0 = \left[ 
\begin{array}{c}
\boldsymbol 0 \\
1\\
\end{array}	
\right], \ \ \ \  
\boldsymbol{\tilde\omega}^R \equiv \left[ 
\begin{array}{c}
\boldsymbol 1 \\
1\\
\end{array}	
\right].
\end{equation}
$\boldsymbol{\tilde \omega}^L$ is chosen as the left boundary vector because the initial hidden vector $\boldsymbol h_0$ fed into vanilla RNNs is typically chosen to be a zero vector, whereas $\boldsymbol{\tilde \omega}^R$ is chosen as the right boundary vector to ensure that all the output components are taken into account. But, in the large $T$ limit, these choices should not significantly change the entanglement entropy when the bipartition is taken at the bond $ \lceil T/2 \rceil$, which are far away from the boundary.

\subsection{RACs with a pre-trained embedding layer}

\begin{figure}
\includegraphics[width = \textwidth]{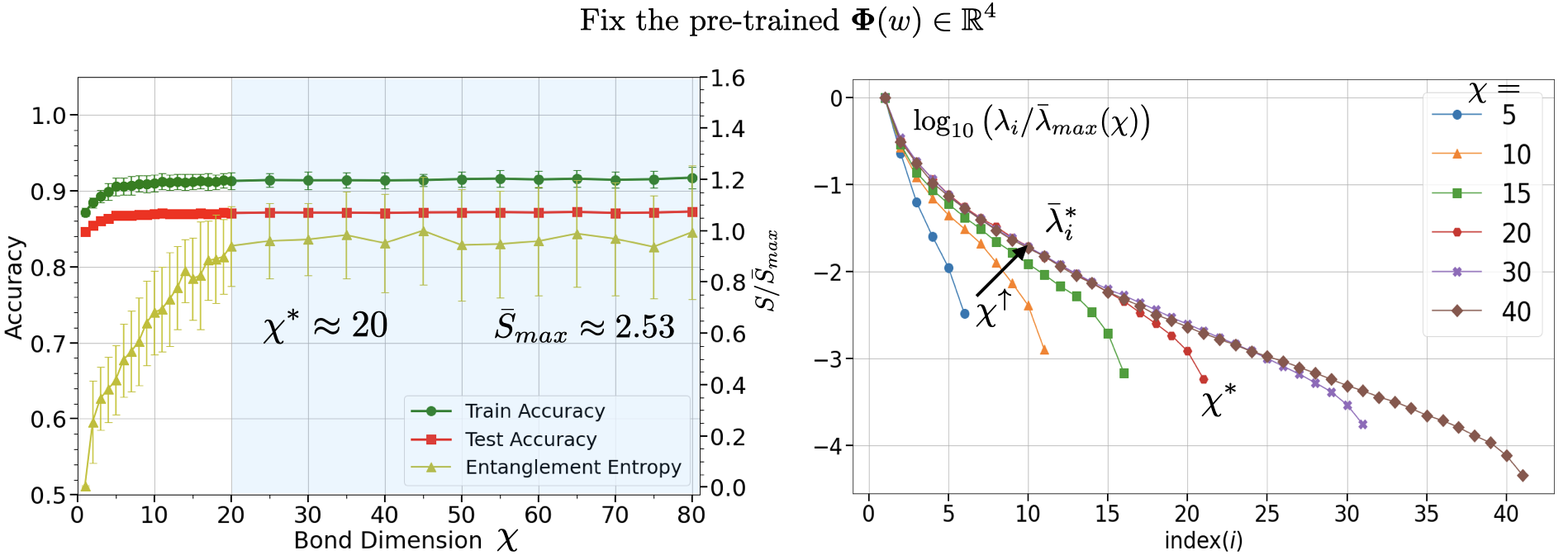}
\caption{The behaviors of trained RACs with additive biases for sentiment analysis of the IMDb data set, when the fixed pre-trained word embedding has dimension $d_I = 4$. (Left) the prediction accuracies saturate when the entanglement entropy (EE) saturates. The critical bond dimension $\chi^* \approx 20$, at which the EE is maximal, enables one to infer a minimal single-layer RACs model for IMDb sentiment analysis. The average maximum entanglement entropy $\bar S_{max}$, which is below the upper bound from the area law, is attained when $\chi \gtrsim \chi^*$. (Right) above the critical bond dimension, the Schmidt coefficients collapse onto the limiting entanglement spectrum $\bar{\lambda^*}_i$ that sets the slowest exponential decay rate of the Schmidt coefficients. Here, the average maximum Schmidt coefficients for $\chi =5,10,15,20,30,40$ are $\bar{\lambda}_{max}(\chi)\approx 0.75, 0.63, 0.58, 0.51, 0.51, 0.52,$ respectively. The average are taken over 50 trained models; each begins with a random initialization of RACs with additive biases. The error bars for $\lambda_i$ are not shown for the clarity of presentation.}
\label{fig: EE-schmidt fix phi_imdb}
\end{figure}
To isolate the interaction between RACs and the embedding layer, we pre-train the word embedding $\Phi$ (recall section \ref{subsec: embedding}) independently from RACs. First, we train $\Phi$ on the IMDb training data with a flatten layer publicly available in Keras, while the output is still the sigmoid function discussed earlier. The flatten layer contains no trainable parameters and as a result the classification accuracy is optimized based solely on the trainable word embedding. After we train the embedding layer for 100 epochs with the early stopping criterion explained in section \ref{subsec:sentiment}, we arrive at a pre-trained $\Phi$ that is not specifically optimized for RACs, thereby isolating the expressiveness that could arise from the interaction between RACs and the embedding layer. After we obtain this pre-trained embedding layer $\Phi$, $\Phi$ is fixed and training optimizes only weights and biases of RACs. 

Figure ~\ref{fig: EE-cosine train phi_imdb} (left) shows the prediction accuracies and the entanglement entropy as a function of the bond dimension $\chi$ for embedding dimension $d_I=4$.  It can be seen that from bond dimension 1 to approximately 20, the training accuracy increases monotonically from $87.1\%$ to $91.5\%$ while the test accuracy increases from $84.7\%$ to $86.3\%$. Both quantities saturate at $\chi \approx 20 \equiv \chi^*$. The entanglement entropy also increases rapidly before the onset of the accuracy saturation, then for $\chi > \chi^*$ it saturates at the (average) maximum value of $\bar{S}_{max} \approx 2.53$. The results suggest a critical model size $\chi^*$ such that RACs expressiveness is maximal. Above this critical size both the prediction accuracies and the entanglement entropy saturate. For practical purposes, this critical size $\chi^*$ is valuable for identifying a minimal single-layer RACs model that can best estimate the statistics of IMDb training data set.

For the IMDb data set, the minimal model size for single-layer RACs with a fixed pre-trained embedding with $d_I=4$ is $\chi^* \approx 20.$ We also observe similar behaviors on the saturation of prediction accuracies that correspond to the  saturation of entanglement entropy for larger pre-trained embeddings with $d_I = 8, 16, 32$ with the average maximum entanglement entropy of $\bar{S}_{max} \approx 3.87, 4.86, 5.09$, respectively. For larger embedding dimensions, not only the maximum entanglement entropy increases, the critical bond dimensions and the saturated prediction accuracies also increase (not shown here due to redundancy of the plots.) These results are not specific to the IMDb data set, as we observe similar trends in single-layer RACs with a fixed pre-trained embedding in a smaller RT movie review data set as well. The results for the RT data set is provided in the Appendix\footnote{For the embedding dimension $4$, the critical bond dimension for RT movie data set is $\chi^* \approx 40$, beyond which the entanglement entropy very slowly increases and plateaus out at the maximum value of 1.}.

To understand how the entanglement entropy becomes saturated above a critical bond dimension $\chi^*$, we investigate the behaviors of the average Schmidt coefficients for the model size from $\chi=5$ to $\chi=40$. Interestingly, figure \ref{fig: EE-cosine train phi_imdb}(right) reveals that above the critical model size, the larger values of the entanglement spectrum (the function defined by the Schmidt coefficients indexed in a descending order)  all collapse onto a {\it limiting entanglement spectrum} $\bar \lambda^*_i$, which exhibits the {\it slowest possible} exponential decay rate of the Schmidt coefficients.Thus this limiting entanglement spectrum $\bar \lambda^*_i$ defines the average maximum entanglement entropy achievable by our MPS ansatz for this data set, whose value is given by
\begin{equation}\label{eqn: MaxEE}
	\bar{S}_{max}=-\sum_{i=1}^{\chi^*} \bar{\lambda^*}_{i}^{2} \log _{2} \bar{\lambda^*}_{i}^{2}.
\end{equation} 
This unique explainability of RACs allows us to infer a minimal recurrent neural networks-based model with the minimal number of hidden neurons $d_H^* + 1 = \chi^*$ for a given task, which is not possible with standard RNNs. From statistical learning theory point of view, the limiting function $\bar{\lambda^*}_i$ determines the bias (in the bias-variance tradeoff sense) of single-layer RACs, which constrains the information propagation capacity as measured by the average maximum entanglement entropy \eqref{eqn: MaxEE}. It is interesting to note that the maximum entanglement entropy is below the upper bound from the area law of $\log_2(\chi)$, as $\bar S_{max} \approx 1.17 < 4.32 \approx \log_2(\chi = 20).$ Hence, a realistic sequence modeling task such as sentiment analysis can still achieve high prediction accuracies using easily trainable RACs, even when the maximum information propagation is bounded above. In fact, the embedding layer $\Phi$ plays a crucial role in attaining high expressive power, as we show next.

\subsection{The interplay between RACs and the word embedding}
\begin{figure}
\includegraphics[width = \textwidth]{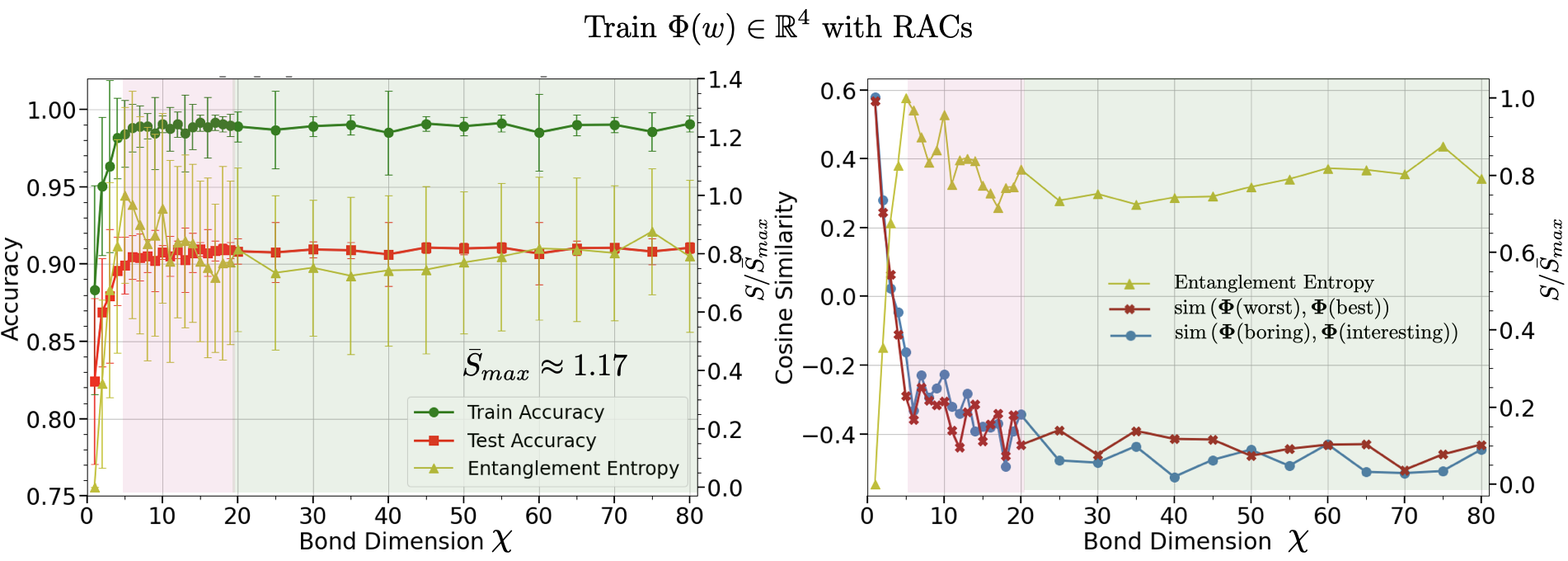}
\caption{The behaviors of RACs with additive biases for sentiment analysis of the IMDb data set, when the word embedding of dimension $d_I =4$ is trained together with RACs. (Left)  the prediction accuracies slowly increase when the entanglement entropy (EE) drops down from the maximum value $\bar S_{max} \approx 1.17$ at $\chi \approx 5$ to the saturated value $ 0.8\bar S_{max}$ at $\chi \approx 20$, after which the accuracies also saturate. The saturated EE here is much smaller than that of the fixed embedding case, suggesting that the expressivity is harnessed from a more meaningful representation $\Phi$. (Right) 
The cosine similarity between the two word embedding computed from \eqref{eq: cosinesim} reveals that indeed the expressivity is boosted via meaningful word embeddings, which arise in larger single-layer RACs.}
\label{fig: EE-cosine train phi_imdb}
\end{figure}
To analyze the interplay between the recurrent units in RACs and the embedding layer $\Phi$, we now train both components simultaneously. The prediction accuracy and the entanglement entropy as a function of the bond dimension is depicted in figure \ref{fig: EE-cosine train phi_imdb}(left). It can be seen that the training and the test accuracy rapidly increases to $98.3 \%$ and $90 \%$, respectively, at $\chi = 5$. The training accuracy then saturates and fluctuates mildly around $98.5\%$ for $\chi > 5$, while the test accuracy slowly increases for $\chi \in [5,20]$, after which it saturates at around $90.5 \%$ accuracy. Despite being simple, our model is ranked 21 (out of 35) in top-performing models (measured by test accuracy) for IMDb sentiment analysis~\cite{papercode_imdb}. The best performing model~\cite{imdb_r1} achieving the test accuracy of $97.2\%$ also uses simple neural network architecture but with the improved quality of the word embeddings. Interestingly, unlike in the fixed word embedding case where the maximum of entanglement entropy (EE)  is attained at its saturation, here the EE attains its maximum at $\chi\approx 5$ at the value of $\bar{S}_{max} \approx 1.17$ before dropping down and saturating at 0.8$\bar{S}_{max}$ when $\chi \approx 20$, after which it fluctuates mildly around the saturated value. 

Although the EE drops after its peak value, the prediction accuracies counter-intuitively increase. Also, compared to the fixed embedding case, $\bar S_{max}$ here is smaller and the prediction accuracies, especially the training accuracy, are higher. These behaviors also arise in larger word embedding size of $d_I = 8,16,32$, though the maximum entanglement entropy at the peak are larger and occurs at a larger bond dimension for a larger model (The plots are not shown here due to redundancy). The larger model also attains higher saturated prediction accuracies. 
These results suggest that information propagation or long-range temporal correlation in sequence modeling is not the main source of expressiveness in estimating the distribution $P(\sigma | w_{1:T})$ in sentiment analysis tasks. In fact, the drop in the EE as the accuracies increase suggests that single-layer RACs must have gained the expressivity through the word embedding $\Phi$.

To test the hypothesis, we plot the cosine similarity \eqref{eq: cosinesim} between embedding vectors of two opposite words that most frequently appear and tends to have a strong influence on the review sentiment, i.e. `boring' and `interesting', `worst' and `best', depicted in figure \ref{fig: EE-cosine train phi_imdb}(right). We see that the cosine similarity drops monotonically with the bond dimension and saturates at $\chi \approx 5$. This might suggest that for $\chi < 5$, the prediction accuracy stems mostly from the temporal correlation in RACs, while at $\chi \ge 5$, the word embedding layer better learns word semantics and start to contribute to higher prediction accuracy.

\section{Discussion and outlook}\label{sec:discussion}

We have recasted single-layer recurrent arithmetic circuits (RACs) with additive biases as the dual matrix product states for the entanglement entropy analysis of a real-world sequence modeling task,
the sentiment analysis of large realistic movie review data sets. The results elucidate that, although the entanglement entropy of the models is bounded above, single-layer RACs can harness their expressive power from trainable word embedding $\boldsymbol \Phi$, achieving considerably high prediction accuracies. Even for a fixed word embedding, single-layer RACs can already achieve high prediction accuracies that saturate when the entanglement entropy saturates at its maximum value $\bar{S}_{max}$. This $\bar{S}_{max}$  allows one to identify the minimal bond dimension $\chi^*$ that RACs can best approximate the sentiment distribution of data sequence $P(\sigma | w_{1:T})$. This $\bar{S}_{max}$ is also {\it below} the upper bound of the area law for entanglement entropy of a matrix product state. Therefore, for sentiment analysis tasks, a low entanglement entropy is not a warrant to disregard simple yet easily trainable models such as single-layer RACs. Importantly,  the crucial interplay between information propagation in the recurrent networks (as reflected by the entanglement entropy) and the meaningful word embedding $\boldsymbol \Phi$ enables single-layer RACs to very well estimate the sentiment distribution of word sequence $P(\sigma | w_{1:T})$. Our analysis also quantitatively reveals the nature of movie review sentiment analysis that NLP practitioners are intuitively aware of; reading only a few statements that contain meaningful keywords might be an efficient strategy to correctly classify the sentiment of a long review. 

Despite the simplicity of our single-layer architecture with low-dimensional word embeddings, we still achieve the test accuracy of $90.5\%$ for the sentiment analysis of the IMDb data set, placing our minimal model in the list of top-performing models~\cite{papercode_imdb}. Some top-performing models utilize powerful modern neural network architectures such as graph neural networks~\cite{gnn_r4} or transformers~\cite{NEURIPS2019_transformer_r2, transformer_r3}. All of which still lack explainability.  It is interesting to note that some simple models, such as classic LSTM architectures (with high quality word embeddings)~\cite{lstm_embedding_1,lstm_embedding_2}, are also in the top-performing list. Remarkably, the best performing model utilizes a very simple neural architecture with the emphasis on constructing highest quality word embeddings~\cite{imdb_r1}. This observation agrees with our quantitative evidence that long-range information propagation is not the main source for RNNs' successes in sentiment analysis, but high model expressiveness can be attained from the subtle interplay between the information propagation and the quality of word vector embeddings.

It would be interesting to generalize the current analysis to deep (multi-layer) RACs models \cite{levine2018benefits} to see the interplay between long-range information propagation in the recurrent networks and the meaningful word embedding in other realistic natural language processing tasks, such as sequence to sequence modeling. Perhaps one could also find a minimal deep RACs model that reproduces the power-law decay in the mutual information between characters, which is a feature of classical English texts \cite{Lin_Tegmark_Entropy, Cirac_2021}. Recently, variants of standard many-body quantum states have been analyzed as highly expressive variational ansatz to estimate probability distribution \cite{Cirac_IEEE20, Miller_Terilla_2021,Glasser_NEURIPS2019}; it'd also be interesting to implement such models for realistic natural language processing tasks and investigate how word embedding could help boost models prediction accuracy. Lastly, regarding the limiting entanglement spectrum that sets the maximum entanglement entropy of single-layer RACs,  theoretical understanding of such entanglement spectrum may hint at the minimum bias (in the bias-variance tradeoff sense) attainable by RACs to estimate a data distribution, which could provide a guideline to systematically study the expressive power of recurrent neural networks from statistical learning theory viewpoints.

\ack
This research has received funding support from the National Science, Research and Innovation Fund (NSRF) via the Program Management Unit for Human Resources \& Institutional Development, Research and Innovation [grant number B05F640051], and from Thailand Science Research and Innovation Fund Chulalongkorn University [CU\_FRB65\_ind (5)\_110\_23\_40]. J.Tangpanitanon, and P. Bhadola are supported by Blueqat Inc. We acknowledge the National Science and Technology Development Agency, National e-Science Infrastructure Consortium, Chulalongkorn University and the Chulalongkorn Academic Advancement into Its 2nd Century Project (Thailand) for providing computing infrastructure that has contributed to the research results reported within this paper (URL:www.e-science.in.th.)  We also thank V. Ngampruetikorn for a useful discussion, A. T. Rutherford and C. Polpanumas for providing helpful feedbacks on the manuscript, and K. Phornsiricharoenphant for providing technical supports on computational hardware used in this work.

\section*{References}
\bibliographystyle{iopart-num.bst}
\bibliography{MPS_RNN_NJP2021_Submission_v2.bib}

\appendix
\newpage 
\section{Sentiment analysis results for the Rotten Tomatoes movie review data set}

To show that our conclusions apply to other data set, here we report similar results for sentiment analysis of the Rotten Tomatoes (RT) movie review data set. RT movie review data set is a smaller standard data set for sentiment analysis using binary classification \cite{PangLee:05a}. The training set and the test set contain $M=$ 8,400 and 2,662 different samples respectively. Both sets are balanced such that each set contains an equal number of positive and negative reviews. The length of each review is set to $T=20$ and the dictionary size is $N=3,000$.  Similar training procedure to those in the main text is applied. The batch size, however, is set to $32$ for this smaller data set. 

\begin{figure}
\includegraphics[width = \textwidth]{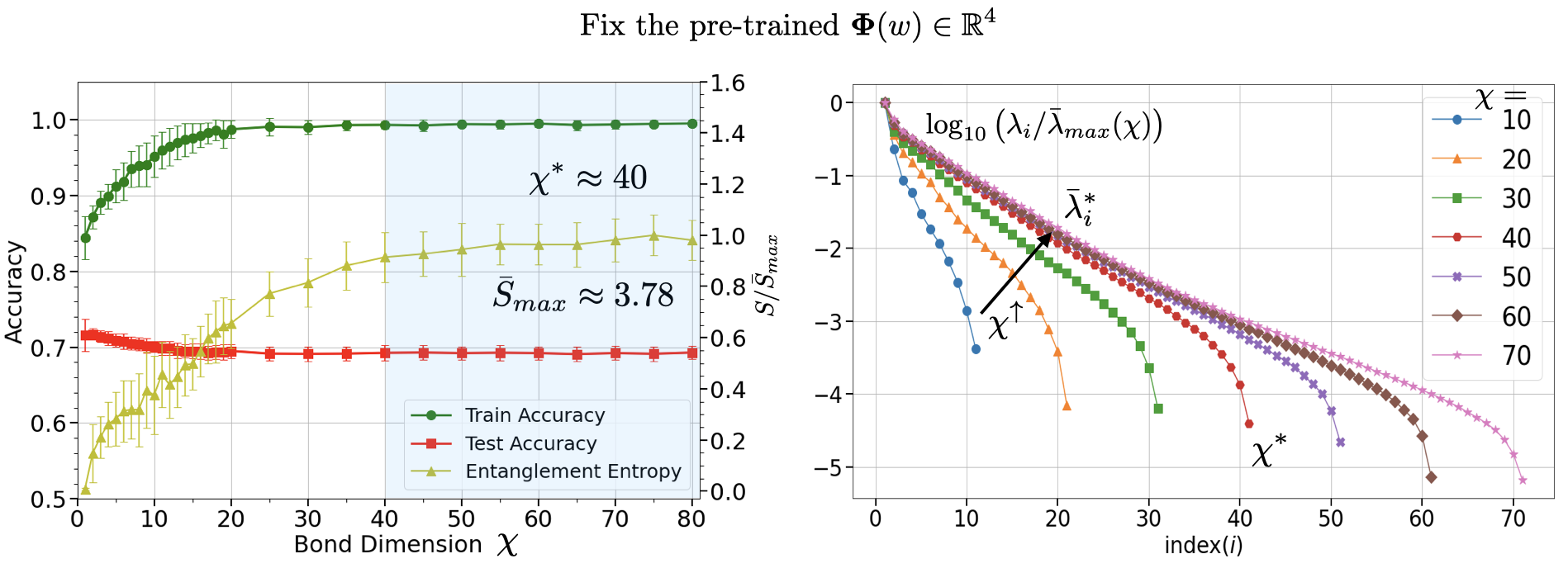}
\caption{The behaviors of trained RACs with additive biases for sentiment analysis of the Rotten Tomatoes data set, when the fixed pre-trained word embedding has dimension $d_I = 4$. (Left) the prediction accuracies saturate when the entanglement entropy (EE) very slowly increases to saturation. The critical bond dimension $\chi^* \approx 40$, after which the EE very slowly increases and eventually saturates, enables one to infer a minimal single-layer RACs model for this data set. The average maximum entanglement entropy $\bar S_{max}$, which is below the upper bound from the area law, is attained when $\chi \gtrsim \chi^*$. (Right) above the critical bond dimension, the Schmidt coefficients almost collapse onto the limiting entanglement spectrum $\bar{\lambda^*}_i$ that sets the slowest exponential decay rate of the Schmidt coefficients. Here, the average maximum Schmidt coefficients for $\chi =10,20,30,40,50,60,70$ are $\bar{\lambda}_{max}(\chi)\approx 0.69, 0.48, 0.37, 0.30, 0.28, 0.27, 0.25,$ respectively. The average are taken over 50 trained models; each begins with a random initialization of RACs with additive biases. The error bars for $\lambda_i$ are not shown for the clarity of presentation.}
\label{fig: EE-schmidt fix phi_rt}
\end{figure}

Figure~\ref{fig: EE-schmidt fix phi_rt}(left) shows the prediction accuracy and the entanglement entropy as a function of the bond dimension $\chi$ for embedding dimension $d_I=4$.  It can be seen that from bond dimension 1 to approximately 40, the training accuracy increases monotonically from $82.2\%$ to $99.3\%$ while the test accuracy drops from $71.8\%$ to $69.1\%$. Both quantities saturate at $\chi \approx 40\equiv \chi^*$. The increase in the training accuracy and the decrease in the test accuracy as the number of model parameters increases suggests that the model is overfitting, which can perhaps be alleviated by adding Dropout though it's not clear whether RACs with Dropout can be mapped to MPS. On the other hand, the entanglement entropy increases rapidly before the onset of the prediction accuracy saturation at $\chi^*$, beyond which it almost plateaus out at large $\chi$. The results suggest a critical model size $\chi^*$ such that RACs expressiveness is maximal. Above this critical size the prediction accuracies saturate, and the entanglement entropy increases very slowly or plateaus out. Similar to the IMDb data set, this critical size $\chi^*$ is valuable for identifying a minimal model that can achieve highest training accuracies for this class of model architecture.

For RT data set, the minimal model size for single-layer RACs with a fixed pre-trained embedding with $d_I=4$ is $\chi^* \approx 40.$ We also observe similar behaviors on the saturation of prediction accuracies that correspond to the  saturation of entanglement entropy for larger pre-trained embeddings with $d_I = 8, 16, 32$ with the average maximum entanglement entropy of $\bar{S}_{max} \approx 3.90, 4.14, 4.38$, respectively. For larger embedding dimensions, not only the maximum entanglement entropy increases, the critical bond dimensions and the saturated prediction accuracies are also larger (not shown here due to redundancy of the plots.)

 Fig. \ref{fig: EE-schmidt fix phi_rt}(right) reveals that above the critical model size $\chi^*$, the Schmidt coefficients (indexed in a descending order) are converging towards the limiting $\bar \lambda^*_i$, which, similar to the IMDb data set in the main text, constrains the {\it slowest possible} exponential decay rate of the Schmidt coefficients. This limiting entanglement spectrum $\bar \lambda^*_i$  should constrain the average maximum entanglement entropy according to \eqref{eqn: MaxEE} and also defines the  bias (in the bias-variance tradeoff sense) in the RACs architecture for sentiment analysis modeling. Similar to the IMDb data set, we also note that the maximum entanglement entropy is below the upper bound from the area law of $\log_2(\chi)$, as $\bar S_{max} \approx 1.20 < 5.32 \approx \log_2(\chi = 40).$ 
 
\begin{figure}
\includegraphics[width = \textwidth]{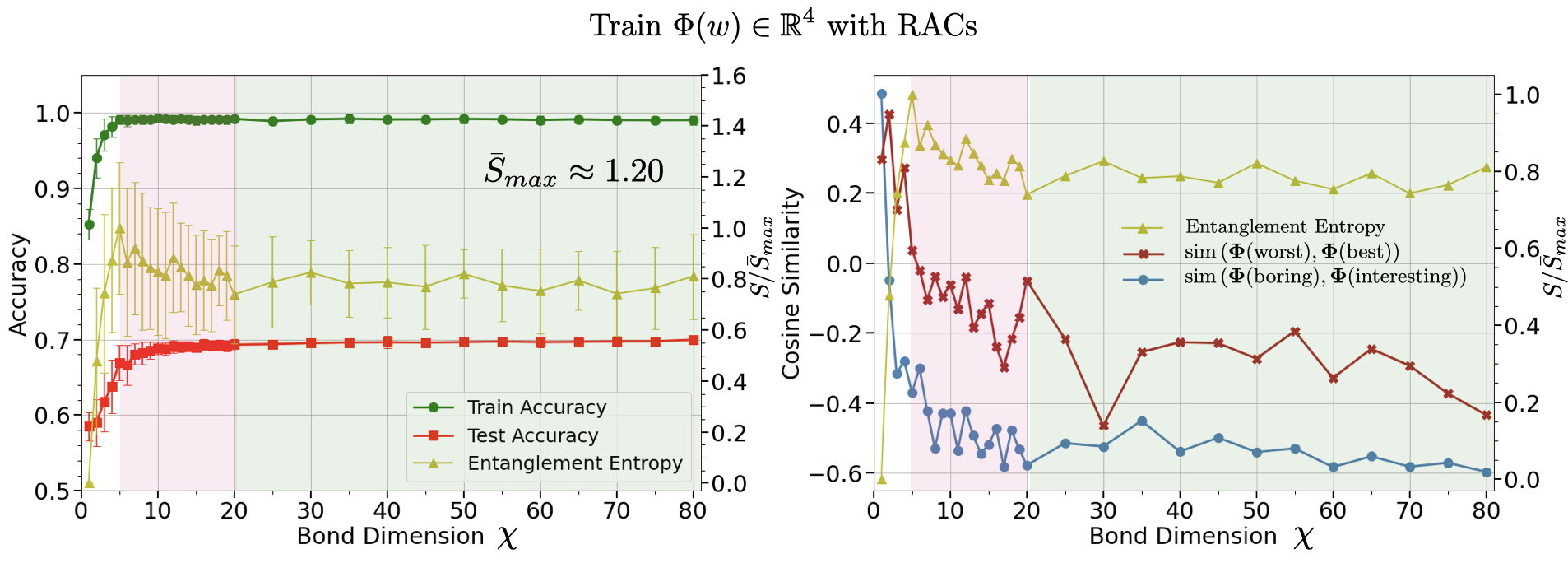}
\caption{The behaviors of RACs with additive biases for sentiment analysis of the Rotten Tomatoes data set, when the word embedding of dimension $d_I =4$ is trained together with RACs. (Left)  the prediction accuracies slowly increase when the entanglement entropy (EE) drops down from the maximum value $\bar S_{max} \approx 1.20$ at $\chi \approx 5$ to the saturated value $ 0.8\bar S_{max}$ at $\chi \approx 20$, after which the accuracies also saturate. The saturated EE here is much smaller than that of the fixed embedding case, suggesting that the expressivity is harnessed from a more meaningful representation $\Phi$. (Right) 
The cosine similarity between the two word embedding computed from \eqref{eq: cosinesim} reveals that indeed the expressivity is boosted via meaningful word embeddings, which arise in larger single-layer RACs.}
\label{fig: EE-cosine train phi_rt}
\end{figure}

To analyze the interplay between RACs and the embedding layer, we now train both components simultaneously. The prediction accuracy and the entanglement entropy as a function of the bond dimension is depicted in Fig.\ref{fig: EE-cosine train phi_rt}(left). It can be seen that the training accuracy increases monotonically to saturation with a $99.10 \%$ accuracy at $\chi\approx 5$, while the test accuracy rapidly increases to $67 \%$ at $\chi = 5$ then gradually increases to saturation with a $70 \%$ accuracy at $\chi \approx 20$.  On the other hand, the entanglement entropy displays a peak at $\chi\approx 5$ before dropping rather steadily to saturation at $\approx 0.8\bar S_{max}$ at $\chi \approx 20$, after which it fluctuates mildly around the saturated value. The test accuracy in our simple setting is comparable to the last entry in the state-of-the-art list for Rotten Tomatoes sentiment analysis~\cite{papercode_mr}, which achieves $76\%$ test accuracy using graph convolutional neural network architecture~\cite{mr_r15}. 

Similar to IMDb data set in the main text, the EE drops after its peak value, while the prediction accuracies  increase. These behaviors also arise in larger word embedding sizes, though the maximum entanglement entropy at the peak are larger ($d_I = 8,16,32$,  $\bar{S}_{max} \approx 1.68, 2.08, 2.51,$) and occurs at a larger bond dimension for a larger model (plots are not shown here due to redundancy.)  
The decay in the EE that corresponds to the increase in the prediction accuracies  can be attributed to a more meaningful word embedding $\boldsymbol \Phi$, as shown in the cosine similarity plots between embedding vectors of the two opposite words, depicted in Fig. \ref{fig: EE-cosine train phi_rt}(right).

\end{document}